\documentclass[12pt,a4paper]{article}
\usepackage{epsfig}
\usepackage{amsmath,amsfonts,amssymb}
\usepackage{t1enc}
\usepackage{cite}

\begin{document}
\vspace*{-3cm}
\begin{flushright}
hep-ph/0409342 \\
September 2004
\end{flushright}

\renewcommand{\thefootnote}{\fnsymbol{footnote}}

\begin{center}
\begin{Large}
{\bf Top flavour-changing neutral interactions: \\[0.2cm] 
theoretical expectations and  experimental detection
\footnote{Presented at the final meeting of the European Network ``Physics at
Colliders'', Montpellier, September 26-27, 2004.}
}
\end{Large}

\vspace{0.5cm}
J. A. Aguilar--Saavedra \\[0.2cm]
{\it Departamento de Física and CFTP, \\
  Instituto Superior Técnico, P-1049-001 Lisboa, Portugal} \\
\end{center}

\begin{abstract}
Top flavour-changing neutral interactions with a light quark $q=u,c$ and a gauge
or Higgs boson are very 
suppressed within the Standard Model (SM), but can reach observable levels in 
many of its extensions. We review the possible size of the effective vertices 
$Ztq$, $\gamma tq$, $gtq$ and $Htq$ in several SM extensions, and discuss the 
processes in which these interactions might show up at LHC and at a high energy
$e^+ e^-$ linear collider.
\end{abstract}

\setcounter{footnote}{0}
\renewcommand{\thefootnote}{\arabic{footnote}}

\section{Introduction}

The next generation of high energy colliders planned or under construction will
test the Standard Model (SM) with high precision and will explore higher
energies in
the search of new physics. New physics may manifest itself in two ways: through
direct signals involving the production of new particles or
by departures from the SM predictions for the known
particles. Direct signals are crucial in order to establish the type of
new physics present in nature but indirect effects are important as well, and
in some cases they could give evidence of physics beyond the SM before new
particles are discovered.

The top quark plays a key role in the quest for deviations from SM predictions
for two reasons: ({\em i\/}) due to its large mass, radiative corrections
involving new particles are often more important than for
lighter fermions; ({\em ii\/}) its large mass suggests that it might
have a special role in electroweak symmetry breaking. Top quarks will be
copiously produced at LHC and, to a lesser extent, at a high energy $e^+
e^-$ collider like TESLA. With such large samples, precise measurements of
its couplings will be available to test SM
predictions \cite{top,tesla}. Here we study flavour-changing
neutral (FCN)
couplings involving the top quark. The most general effective Lagrangian
describing its interactions with a light quark $q=u,c$ and a gauge or Higgs
boson, containing terms up to dimension 5, can be written as
\begin{eqnarray}
-\mathcal{L}^\mathrm{eff} & = & \frac{g}{2 c_W} X_{qt} \, \bar q \gamma_\mu
(x_{qt}^L P_L + x_{qt}^R P_R) t Z^\mu 
+ \frac{g}{2 c_W} \kappa_{qt} \, \bar q (\kappa_{qt}^v +\kappa_{qt}^a \gamma_5)
\frac{i \sigma_{\mu \nu} q^\nu}{m_t} t Z^\mu  \nonumber \\
& &  + e \lambda_{qt} \, \bar q (\lambda_{qt}^v + \lambda_{qt}^a \gamma_5)
\frac{i \sigma_{\mu \nu} q^\nu}{m_t} t A^\mu
+ g_s \zeta_{qt} \, \bar q (\zeta_{tq}^v + \zeta_{qt}^a \gamma_5)
\frac{i \sigma_{\mu \nu} q^\nu}{m_t} T^a q G^{a\mu} \nonumber \\
& & + \frac{g}{2 \sqrt 2} g_{qt} \, \bar q (g_{qt}^v + g_{qt}^a \gamma_5) t H
+ \mathrm{H.c.} \,,
\label{ec:1}
\end{eqnarray}
where $q^\nu = (p_t-p_q)^\nu$ is the boson momentum and $\bar q$, $t$ are
shorthands for the quark fields $\bar u(p_q)$ and $u(p_t)$, respectively. The
couplings are constants
corresponding to the first terms in the expansion in momenta,
normalised to $|x_{qt}^L|^2+|x_{qt}^R|^2=1$,
$|\kappa_{qt}^v|^2+|\kappa_{qt}^a|^2=1$, etc., with $X_{qt}$,
$\kappa_{qt}$, $\lambda_{qt}$, $\zeta_{qt}$ and $g_{qt}$ real and positive.
In principle there are
additional terms that could be included in this effective Lagrangian, for
instance proportional to
$\sigma_{\mu \nu} (p_t+p_q)^\nu Z^\mu$. However, in the processes discussed
the top quark
can be considered on its mass shell to a very good approximation and the gauge
bosons are either on their mass shell or coupling to light fermions. Hence,
these extra interactions can be rewritten in terms of the ones in
Eq.~(\ref{ec:1}) using Gordon identities.

Within the SM, the $\gamma_\mu$ couplings $x_{qt}^{L,R}$ vanish at the tree
level by the GIM mechanism, and non-renormalisable $\sigma_{\mu \nu}$ terms 
do not appear in the Lagrangian. Both types of vertices are generated at one
loop level but, as will be shown in Section \ref{sec:2}, they are strongly
suppressed by the GIM mechanism, making FCN top interactions very
small. In
models beyond the SM this GIM suppression can be relaxed, and
one-loop diagrams mediated by new
bosons may also contribute, yielding effective
couplings orders of magnitude larger than those of the SM.
The possible size of top FCN vertices in several SM extensions will be
summarised in Section \ref{sec:3}. These interactions
lead to various top decay and single top production processes which will be
discussed in Section \ref{sec:4}. The
observation of such processes, extremely rare in the SM,
would provide a clear indirect signal of new physics, although the presence of
SM backgrounds must be considered. In specific models, the presence of
these interactions may be correlated with other effects at high or low energies.
One example of such correlation will be shown in Section \ref{sec:5}.

We note that in the literature there are numerous alternative normalisations of
the coupling constants in
$\mathcal{L}^\mathrm{eff}$. For this
reason, we express our limits on the couplings in terms of top decay
branching ratios. We use $m_t = 178.0 \pm 4.3$ GeV \cite{tmass},
$\alpha(m_t) = 1/128.921$,
$s_W^2(m_t) = 0.2342$, $\alpha_s(m_t) = 0.108$ and assume $m_H = 115$ GeV. 
The tree-level prediction for the leading decay mode $t \to bW^+$ is
\begin{equation}
\Gamma(t \to bW^+) = \frac{\alpha}{16 \, s_W^2} |V_{tb}|^2 \frac{m_t^3}{M_W^2}
\left[ 1-3 \frac{M_W^4}{m_t^4} + 2 \frac{M_W^6}{m_t^6} \right] \,,
\label{ec:2}
\end{equation}
which yields $\Gamma(t \to b W^+) = 1.61$ GeV. We take this value as the total
top width $\Gamma_t$. The partial widths for FCN decays are given by
\begin{eqnarray}
\Gamma(t \to qZ)_\gamma & = & \frac{\alpha}{32 \, s_W^2 c_W^2} |X_{qt}|^2 \,
  \frac{m_t^3}{M_Z^2} \left[ 1-\frac{M_Z^2}{m_t^2} \right]^2
  \left[ 1+2 \frac{M_Z^2}{m_t^2} \right] \,, \nonumber \\
\Gamma(t \to qZ)_\sigma & = & \frac{\alpha}{16 \, s_W^2 c_W^2} 
\left| \kappa_{qt} \right|^2 m_t
\left[ 1-\frac{M_Z^2}{m_t^2} \right]^2 \left[ 2+\frac{M_Z^2}{m_t^2} \right]
 \,, \nonumber \\
\Gamma(t \to q \gamma) & = & \frac{\alpha}{2} 
\left| \lambda_{qt} \right|^2 m_t \,, \nonumber \\
\Gamma(t \to q g) & = & \frac{2 \alpha_s}{3} 
\left| \zeta_{qt} \right|^2 m_t \,, \nonumber \\
\Gamma(t \to q H) & = & \frac{\alpha}{32 \, s_W^2} |g_{qt}|^2 \, m_t
\left[ 1-\frac{M_H^2}{m_t^2} \right]^2 \,.
\label{ec:3}
\end{eqnarray}
The corresponding branching ratios are then
\begin{align}
& \mathrm{Br}(t \to qZ)_\gamma = 0.472 \; X_{qt}^2 \,, \nonumber \\
& \mathrm{Br}(t \to qZ)_\sigma = 0.367 \; \kappa_{qt}^2 \,, \nonumber \\
& \mathrm{Br}(t \to q \gamma) = 0.428 \; \lambda_{qt}^2 \,, \nonumber \\
& \mathrm{Br}(t \to qg) = 7.93 \; \zeta_{qt}^2 \,, \nonumber \\
& \mathrm{Br}(t \to qH) = 3.88 \times 10^{-2} \; g_{qt}^2 \,.
\label{ec:br}
\end{align}

\section{Top FCN interactions in the SM}
\label{sec:2}

\begin{figure}[htb]
\begin{center}
\begin{tabular}{ccc}
\mbox{\epsfig{file=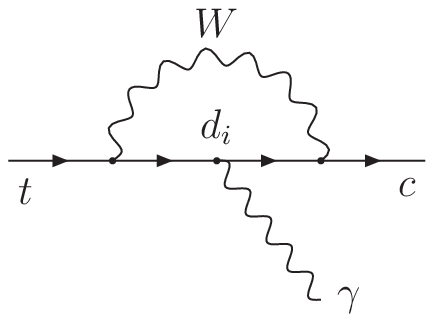,width=3.5cm,clip=}} & ~ &
\mbox{\epsfig{file=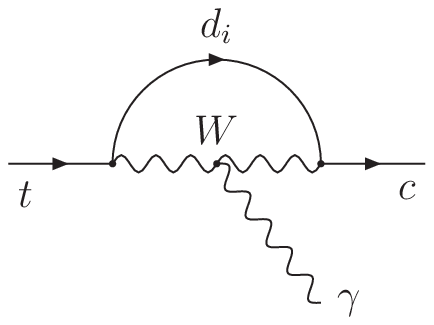,width=3.5cm,clip=}}
\end{tabular}
\end{center}
\caption{SM diagrams contributing to the $tc\gamma$ vertex. The additional
diagrams involving unphysical scalars are not displayed.}
\label{fig:diags}
\end{figure}

One-loop induced FCN couplings involving the top quark have a strong GIM
suppression, resulting in negligible branching ratios for top
FCN decays \cite{hewett,mele}. We
show how this cancellation mechanism operates taking as example
the $\gamma tc$ vertex. The SM diagrams contributing at one loop level are
depicted in Fig.~\ref{fig:diags}, with $d_i=d,s,b$. We omit the diagrams
involving unphysical scalars, which can be obtained replacing the $W$ boson
lines by charged scalars.

\begin{figure}[htb]
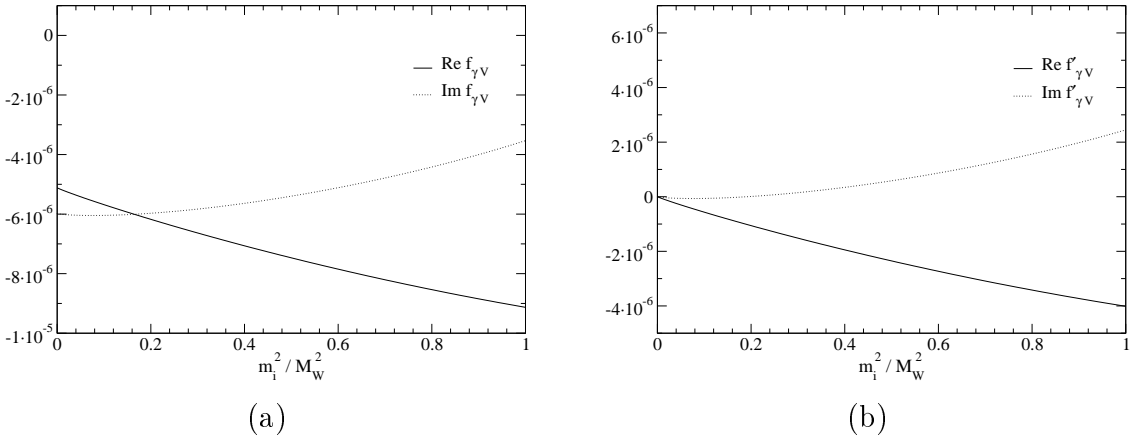

\begin{center}
\begin{tabular}{ccc}
\mbox{\epsfig{file=Figs/A.eps,width=7cm,clip=}} & ~ &
\mbox{\epsfig{file=Figs/Ap.eps,width=7cm,clip=}} \\
(a) & & (b)
\end{tabular}
\end{center}
\caption{Loop functions $f_{\gamma V}(m_i^2/M_W^2)$ and $f'_{\gamma
V}(m_i^2/M_W^2)$.}
\label{fig:f}
\end{figure}

If we define $\mathcal{V}_\gamma \equiv e \lambda_{qt} \lambda_{qt}^v / m_t$,
$\mathcal{A}_\gamma \equiv e \lambda_{qt} \lambda_{qt}^a / m_t$, we can
write these form factors as
\begin{eqnarray}
\mathcal{V}_\gamma & = & \sum_{i=1}^3 f_{\gamma V}(m_i^2/M_W^2) V_{ci} V_{ti}^*
 \,, \nonumber \\
\mathcal{A}_\gamma & = & \sum_{i=1}^3 f_{\gamma A}(m_i^2/M_W^2) V_{ci} V_{ti}^*
 \,,
\label{ec:5}
\end{eqnarray}
with $f_{\gamma V}(x) \simeq f_{\gamma A}(x)$ (equal in the limit $m_c=0$)
and $V$ the Cabibbo-Kobayashi-Maskawa (CKM) matrix. The function
$f_{\gamma V}$ is shown in Fig.~\ref{fig:f} (a). Using the fact that
$m_{d,s} \simeq 0$ to an excellent
approximation, the $3 \times 3$ CKM unitarity relation
$V_{cd} V_{td}^* + V_{cs} V_{ts}^* + V_{cb} V_{tb}^* = 0$ implies
\begin{equation}
\mathcal{V}_\gamma = \left[ f_{\gamma V}(m_b^2/M_W^2)-f_{\gamma V}(0) \right]
 V_{cb} V_{tb}^* \equiv f'_{\gamma V}(m_b^2/M_W^2) V_{cb} V_{tb}^*  \,.
\label{ec:6}
\end{equation}
Hence, the form factor is controlled by the shifted function $f'_{\gamma V}$,
plotted in Fig.~\ref{fig:f} (b). We observe that the consequence of $3 \times 3$
CKM unitarity is to cancel
the constant term $f_{\gamma V}(0) \simeq -5.1 \times 10^{-6} - 6.0 \times
10^{-6} i$, common to the three $d,s,b$ contributions, leaving
$\mathcal{V}_\gamma$ proportional to the much smaller function
$f'_{\gamma V}(m_b^2/M_W^2) \simeq f'_{\gamma V}(0.0012)
\simeq -9.1  \times 10^{-9} -4.7 \times 10^{-9} i$.

This cancellation makes the form factors rather sensitive to the value
of the $b$ quark mass in the internal propagators. The most
adequate choice is the running $\overline{\mathrm{MS}}$ mass evaluated
at a scale $O(m_t)$. With $\overline{m_b}(m_t) = 2.74 \pm 0.17$ GeV, the SM
prediction for $t \to c \gamma$ is \cite{bruno}
\begin{equation}
\mathrm{Br}(t \to c \gamma) = (4.6 ~^{+1.2}_{-1.0} \pm 0.2
\pm 0.4 ~^{+1.6}_{-0.5}) \times 10^{-14} \,.
\label{ec:7}
\end{equation}
The first and second uncertainties quoted come from the bottom and top masses,
respectively, the third from CKM matrix elements and the fourth is estimated
varying the renormalisation scale
between $M_Z$ (plus sign) and $1.5 \, m_t$ (minus sign). The
analogous calculation of $t \to cg$ yields
\begin{equation}
\mathrm{Br}(t \to c g) = (4.6  ~^{+1.1}_{-0.9} \pm 0.2
 \pm 0.4 ~^{+2.1}_{-0.7}) \times 10^{-12} \,.
\label{ec:8}
\end{equation}
These updated results are
one order of magnitude smaller than the values previously obtained in
Ref.~\cite{hewett}. For $t \to cZ$, $t \to cH$ the results of
Refs.~ \cite{hewett,mele} must be rescaled by a factor
$[\overline{m_b}(m_t)/(5 ~\mathrm{GeV})]^4
\simeq 0.09$ (the loop functions are approximately linear for $m_b^2/M_W^2 \ll
1$), obtaining
\begin{eqnarray}
\mathrm{Br}(t \to cZ) & \simeq & 1 \times 10^{-14} \,, \nonumber \\[0.2cm]
\mathrm{Br}(t \to cH) & \simeq & 3 \times 10^{-15} \,.
\end{eqnarray}
The relative uncertainties on these values are expected to be similar to the
ones in Eqs.~(\ref{ec:7}),(\ref{ec:8}).
For decays $t \to uZ$, $t \to u \gamma$, $t \to ug$, $t \to uH$ the branching
ratios are a factor $|V_{ub}/V_{cb}|^2 \simeq 0.0079$ smaller to the ones
corresponding to a $c$ quark, as can be seen from Eq.~(\ref{ec:6}). The
difference between the $u$ and $c$ masses is irrelevant.

\section{Top FCN interactions beyond the SM}
\label{sec:3}

New physics contributions to the effective Lagrangian in Eq.~(\ref{ec:1})
can enhance the rates of top FCN decays several orders of
magnitude, giving observable branching ratios in some regions
of parameter space. Here we examine the situation in the context of models with
extra quark singlets, with an extra Higgs doublet and in supersymmetric
extensions of the SM.

In models with extra quarks the $3 \times 3$ CKM matrix is no longer unitary and
the GIM mechanism acting to suppress the SM amplitudes is relaxed. When the new
quarks are $\mathrm{SU}(2)_L$ singlets with charge $Q=2/3$, the couplings of the $Z$
boson to up-type quarks are not diagonal. Taking a
conservative value for the mass of the new quark, $m_T \geq 300$ GeV, present
experimental data allow
\begin{equation}
X_{qt} \simeq 0.015  \quad \quad  (|x_{qt}^L| = 1 \,, x_{qt}^R = 0)
\end{equation}
at the tree level \cite{largo}.
Such couplings are possible both for up and charm quarks, but not
simultaneously. In these models there also
exist tree-level FCN scalar interactions, given by
\begin{equation}
g_{qt} \simeq \frac{m_t}{M_W} X_{qt}  \quad \quad (g_{qt}^v = g_{qt}^a) \,.
\end{equation}
The branching ratios for top decays mediated by these vertices are
$\mathrm{Br}(t \to qZ) \simeq 1.1 \times 10^{-4}$, 
$\mathrm{Br}(t \to qH) \simeq 4.1 \times 10^{-5}$, respectively. 
The decay rates for $t \to q \gamma$, $t \to q g$ are also enhanced due to the
partial breaking of $3 \times 3$ CKM unitarity and the presence of extra Feynman
diagrams like those in Fig.~\ref{fig:diags} (a) but with an $u$ or $t$ internal
quark and a $Z$ boson. The rates obtained are
$\mathrm{Br}(t \to q \gamma) \simeq 7.5 \times 10^{-9}$,
$\mathrm{Br}(t \to q g) \simeq 1.5 \times 10^{-7}$ for $X_{qt} \simeq 0.015$.
In models with $Q=-1/3$ singlets the branching ratios are much smaller
\cite{bruno} since CKM unitarity breaking is very constrained by experimental
data. In SM extensions with $\mathrm{SU}(2)_L$ doublets there may also exist
right-handed tree-level FCN couplings $X_{qt}$ \cite{prl}.

FCN interactions with scalars are also present at the tree level in two Higgs
doublet models (2HDMs), unless a discrete symmetry is imposed to forbid them.
The couplings are often assumed to scale with quark masses \cite{cheng},
\begin{equation}
g_{qt} \simeq \frac{\sqrt{m_q m_t}}{M_W}
\end{equation}
up to a factor of order unity, {\em i.e.} $g_{ct} \simeq 0.20$,
$g_{ut} \simeq 0.012$, leading to
$\mathrm{Br}(t \to c H) \simeq 1.5 \times 10^{-3}$,
$\mathrm{Br}(t \to u H) \simeq 5.5 \times 10^{-6}$, respectively.
The new scalar fields also give radiative contributions to the $Ztq$,
$\gamma tq$ and $gtq$ vertices, with diagrams analogous to those
in Fig.~\ref{fig:diags}, replacing the $W$ boson by a charged scalar, and
additional diagrams with an up-type internal quark and a neutral scalar.
The resulting branching ratios can be up to
$\mathrm{Br}(t \to c Z) \sim 10^{-7}$,
$\mathrm{Br}(t \to c \gamma) \sim 10^{-6}$,
$\mathrm{Br}(t \to c g) \sim 10^{-4}$ \cite{luke,atwood}, with smaller values
for decays to an up quark.
In 2HDMs without tree-level scalar FCN couplings, 
charged and neutral Higgs contributions to $\mathcal{L}^\mathrm{eff}$ can still
increase significantly the rates for top FCN decays with respect to the SM
predictions. The maximum values reached are of
the order $\mathrm{Br}(t \to c Z) \sim 10^{-10}$,
$\mathrm{Br}(t \to c \gamma) \sim 10^{-9}$,
$\mathrm{Br}(t \to c g) \sim 10^{-8}$,
$\mathrm{Br}(t \to c H) \sim 10^{-5}$ \cite{atwood,sola1}.

Recent calculations in the context of the Minimal Supersymmetric Standard
Model (MSSM) show that for non-universal squark mass terms
$\mathrm{Br}(t \to q Z) \simeq 2 \times 10^{-6}$,
$\mathrm{Br}(t \to q \gamma) \simeq 2 \times 10^{-6}$,
$\mathrm{Br}(t \to q g) \simeq 10^{-4}$ can be reached while keeping agreement
with low energy data  \cite{li,delepine}.
These results are larger than previous estimates \cite{li2,luca,lopez}.
The branching ratio of $t \to qH$ can be up to
$\mathrm{Br}(t \to q H) \sim 10^{-5}$ \cite{sola2}, assuming squark masses above
200 GeV. In all these decays
the largest contributions to the amplitudes come from gluino exchange
diagrams. In non-minimal supersymmetric models with $R$ parity violation,
top FCN decays can also proceed through baryon number violating interactions,
yielding
$\mathrm{Br}(t \to q Z) \simeq 3 \times 10^{-5}$,
$\mathrm{Br}(t \to q \gamma) \simeq 1 \times 10^{-6}$,
$\mathrm{Br}(t \to q g) \simeq 2 \times 10^{-4}$ \cite{yang1},
$\mathrm{Br}(t \to q H) \sim 10^{-6}$ \cite{yang2}.
(We obtain these values taking $\Lambda = 1$ in Refs.~\cite{yang1,yang2}.)

We collect the data presented in this section in Table~\ref{tab:br}, together
with SM predictions. Two
conclusions can be extracted from these figures: ({\em i\/}) Models with
tree-level FCN couplings to $Z$, $H$ give the largest rates for
decays to these particles, as it is expected; ({\em ii\/}) the radiative
decays $t \to q\gamma$,
$t \to q g$ have largest branching ratios in supersymmetric extensions of the
SM.

\begin{table}[htb]
\vspace*{0.2cm}
\begin{center}
\begin{small}
\begin{tabular}{lcccccc}
   & SM & QS & 2HDM & FC 2HDM & MSSM & $R \!\!\!\!\!\!  \not \quad$ SUSY
   \\
\hline  \\[-0.5cm]
$t \to u Z$ & $8 \times 10^{-17}$ & $1.1 \times 10^{-4}$ 
  & $-$ & $-$
  & $2 \times 10^{-6}$ & $3 \times 10^{-5}$ \\
$t \to u \gamma$ & $3.7 \times 10^{-16}$ & $7.5 \times 10^{-9}$ 
 & $-$ & $-$
 & $2 \times  10^{-6}$ & $1 \times 10^{-6}$ \\
$t \to u g$ & $3.7 \times 10^{-14}$ & $1.5 \times 10^{-7}$ 
  & $-$ & $-$
  & $8 \times 10^{-5}$ & $2 \times 10^{-4}$ \\
$t \to u H$ & $2 \times 10^{-17}$ & $4.1 \times 10^{-5}$ 
  & $5.5 \times 10^{-6}$ & $-$
  & $10^{-5}$ & $\sim 10^{-6}$ \\[0.2cm]
$t \to c Z$ & $1 \times 10^{-14}$ & $1.1 \times 10^{-4}$ 
  & $\sim 10^{-7}$ & $\sim 10^{-10}$
  & $2 \times 10^{-6}$ & $3 \times 10^{-5}$ \\
$t \to c \gamma$ & $4.6 \times 10^{-14}$ & $7.5 \times 10^{-9}$ 
 & $\sim 10^{-6}$ & $\sim 10^{-9}$
 & $2 \times 10^{-6}$ & $1 \times 10^{-6}$ \\
$t \to c g$ & $4.6 \times 10^{-12}$ & $1.5 \times 10^{-7}$ 
  & $\sim 10^{-4}$ & $\sim 10^{-8}$
  & $8 \times 10^{-5}$ & $2 \times 10^{-4}$ \\
$t \to c H$ & $3 \times 10^{-15}$ & $4.1 \times 10^{-5}$ 
  & $1.5 \times 10^{-3}$ & $\sim 10^{-5}$
  & $10^{-5}$ & $\sim 10^{-6}$
\end{tabular}
\end{small}
\end{center}
\caption{Branching ratios for top FCN decays in the SM, models with
$Q=2/3$ quark singlets (QS), a general 2HDM, a flavour-conserving (FC)
2HDM, in the MSSM and with $R$ parity violating SUSY.}
\label{tab:br}
\end{table}

\section{Experimental observation}
\label{sec:4}

Present experimental limits on top FCN couplings come from the non-observation
of the decays $t \to qZ$, $t \to q \gamma$ at Tevatron and the absence of single
top production $e^+ e^- \to t \bar q$ at LEP and $eu \to et$ at HERA. The best
limits are
$\mathrm{Br} (t \to qZ) \leq 0.159$ \cite{opal},
$\mathrm{Br}(t \to q \gamma) \leq 0.032$ \cite{cdf},
$\mathrm{Br}(t \to u \gamma) \leq 0.011$ \cite{zeus,h1} with a 95\% confidence
level (CL), very weak if compared to the
rates which can be achieved in the SM and its extensions.\footnote{Our LEP bound
on $\mathrm{Br} (t \to qZ)$ slightly differs from the one quoted in
Ref.~\cite{opal} because we normalise the rates to $\Gamma(t \to bW^+)$. The
translation into limits on $X_{qt}$ is also different from theirs, because they
assume $X_{ut} = X_{ct}$ while we assume only one coupling is different from
zero, thus obtaining more conservative bounds.} These limits will
improve with Tevatron Run II, and will reach the $10^{-5}$ level
at LHC and TESLA (or other future $e^+ e^-$ collider), opening the possibility
of the experimental observation of top FCN interactions.

\subsection{Observation at LHC}

At LHC top quarks are abundantly produced in $t \bar t$ pairs via standard QCD
interactions, with a cross section around 860 pb \cite{top}. The search for
top FCN couplings can be performed looking for processes in which 
the top quark decays via $t \to qZ$ \cite{han1},
$t \to q \gamma$ \cite{han2}, $t \to qg$ \cite{han3}, $t \to qH$ \cite{gustavo},
mediated by the operators in Eq.~(\ref{ec:1}), while the antitop decays
$\bar t \to  W^- \bar b$. The charge conjugate processes, with standard top
decay
and FCN antitop decay, are also included in the analyses but for brevity we
do not refer to them in the following.
Due to the large QCD backgrounds at LHC, the search for signatures of these
processes must be performed in the leptonic channels $W^- \to \ell^- \bar
\nu_\ell$,
with $\ell=e,\nu$ (with a good $\tau$ tagging this channel could be eventually
included as well). In $Z$ and $H$ decays the channels considered are $Z \to
\ell^+  \ell^-$ and $H \to b \bar b$, respectively. $b$ tagging is used in
order to reduce backgrounds.

\begin{figure}[htb]
\begin{center}
\begin{tabular}{ccccc}
\mbox{\epsfig{file=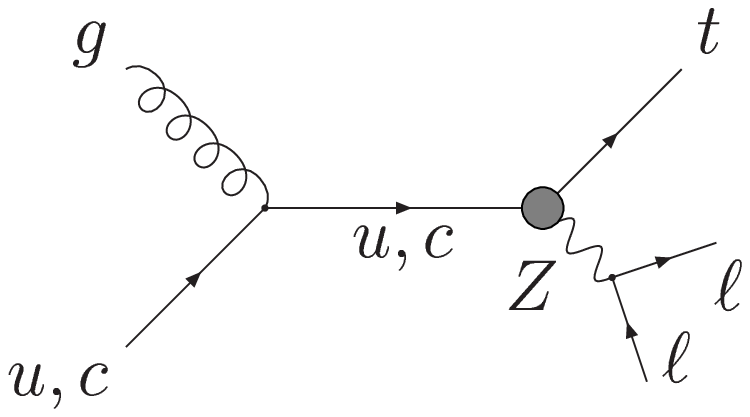,height=1.7cm,clip=}} & 
\mbox{\epsfig{file=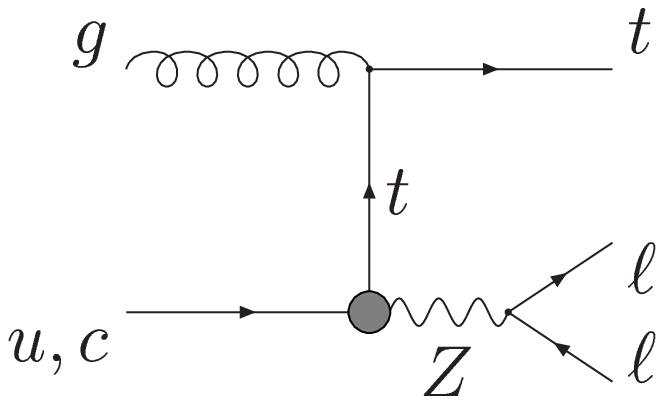,height=1.7cm,clip=}} & ~~ &
\mbox{\epsfig{file=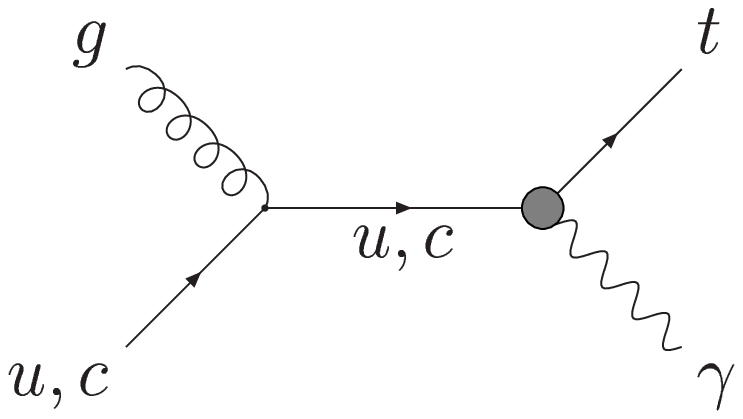,height=1.7cm,clip=}} & 
\mbox{\epsfig{file=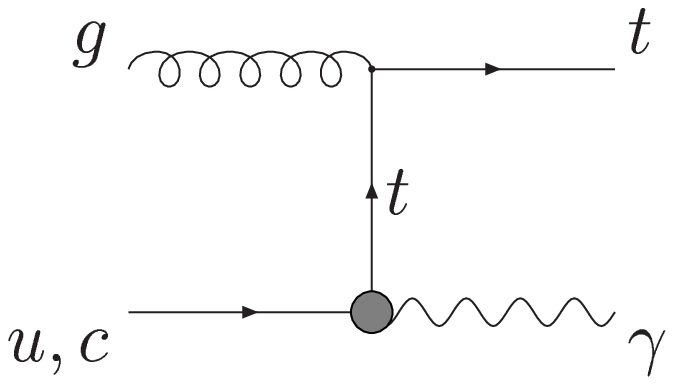,height=1.7cm,clip=}} \\[0.2cm]
\multicolumn{2}{c}{(a)} & & \multicolumn{2}{c}{(b)} \\[0.3cm]
\multicolumn{2}{c}{\epsfig{file=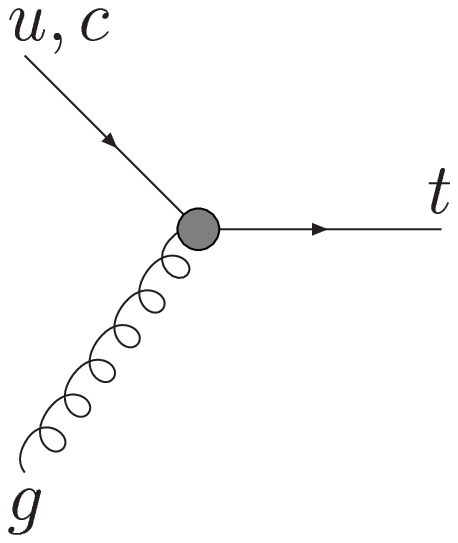,height=1.43cm,clip=}} & &
\mbox{\epsfig{file=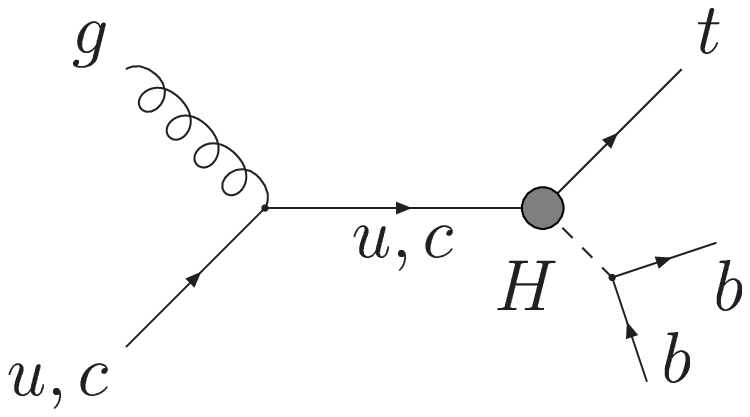,height=1.7cm,clip=}} & 
\mbox{\epsfig{file=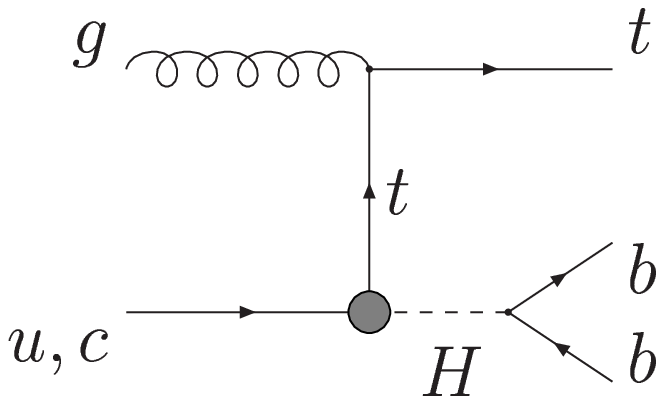,height=1.7cm,clip=}} \\[0.2cm]
\multicolumn{2}{c}{(c)} & & \multicolumn{2}{c}{(d)}
\end{tabular}
\end{center}
\caption{Diagrams for single top production in hadron collisions: (a) $Zt$
production mediated by
$Ztq$ couplings; (b) $\gamma t$ production mediated by $\gamma tq$ couplings;
(c) $t$ production; (d) $Ht$ production.}
\label{fig:single_t}
\end{figure}

On the other hand, one can search for single top production mediated by the
anomalous vertices in Eq.~(\ref{ec:1}), in the processes $gq \to Zt$ \cite{npb},
$gq \to \gamma t$
\cite{npb}, $gq \to t$ \cite{hosch}, $gq \to Ht$ \cite{gustavo}, followed by a
standard top decay $t \to W^+ b$. The Feynman diagrams for these processes are
depicted
in Fig.~\ref{fig:single_t}. $Zt$ and $\gamma t$ production can also occur via
$gtq$ interactions, but the presence of this type of operator is easier to
detect in the process $gq \to t$.
We collect in Table~\ref{tab:cs} the tree-level cross
sections for FCN single top production processes, calculated with MRST
parton distribution functions
set A \cite{mrst}. Next-to-leading order corrections for $Zt$ and $\gamma t$
production
are available for Tevatron energies \cite{kidonakis}. For LHC they are expected
to increase the cross sections at the 10\% level.

\begin{table}[htb]
\vspace*{0.5cm}
\begin{center}
\begin{small}
\begin{tabular}{llcll}
Process & Cross section & & Process & Cross section \\
\hline \\[-0.5cm]
$gu \to Zt$ ($\gamma_\mu$) & $(260 + 50) \, |X_{ut}|^2$ & ~~ &
$gc \to Zt$ ($\gamma_\mu$) & $(26 + 26) \, |X_{ct}|^2$ \\
$gu \to Zt$ ($\sigma_{\mu \nu}$) & $(540 + 87) \, |\kappa_{ut}|^2$ & ~~ &
$gc \to Zt$ ($\sigma_{\mu \nu}$) & $(45 + 45) \, |\kappa_{ct}|^2$ \\
$gu \to \gamma t$ & $(440 + 76) \, |\lambda_{ut}|^2$ & ~~ &
$gc \to \gamma t$ & $(39 + 39) \, |\lambda_{ct}|^2$ \\
$gu \to t$ & $(9.0  + 2.6) \times 10^5 \, |\zeta_{ut}|^2$ & ~~ &
$gc \to t$ & $(1.5 + 1.5) \times 10^5 \, |\zeta_{ct}|^2$ \\
$gu \to H t$ & $(16 + 2.8) \, |g_{ut}|^2$ & ~~ &
$gc \to H t$ & $(1.5 + 1.5) \, |g_{ct}|^2$
\end{tabular}
\end{small}
\end{center}
\caption{Cross sections (in pb) for single top plus antitop production
processes at LHC. In each case the first term in the sum corresponds to the
process quoted and the second term to the charge conjugate process.}
\label{tab:cs}
\end{table}

It is clearly seen that
for $q=c$ these processes are suppressed by the smaller
structure functions for the charm quark. For
nonrenormalisable $\sigma_{\mu \nu}$ couplings the cross sections are
enhanced by the $q^\nu$ factor appearing in the vertex: with the normalisation
chosen for the coupling constants, for $|X_{qt}| \simeq |\kappa_{qt}| \simeq
|\lambda_{qt}|$ the first three branching ratios in Eq.~(\ref{ec:br}) take
similar values, while the cross sections in Table \ref{tab:cs} are much larger
for $\sigma_{\mu \nu}$-type interactions.

The search for these processes is cleaner in the channels where
$W^+ \to \ell^+ \nu_\ell$, $Z \to \ell^+ \ell^-$, $H \to b \bar b$, and taking
advantage of $b$ tagging to reduce
backgrounds. Their experimental signatures are
written in Table~\ref{tab:back}, where we also include the most important
backgrounds. 

\begin{table}[htb]
\begin{center}
\begin{tabular}{lcclccccl}
Process & Signal & \multicolumn{2}{c}{Background} & ~ & Process & Signal &
\multicolumn{2}{c}{Background} \\
\hline \\[-0.5cm]
$t \bar t$, $t \to qZ$ & $\ell^+ \ell^- j \ell \nu b$ & $ZWjj$ & LO & 
 & $gq \to Zt$ & $\ell^+ \ell^- \ell \nu b$ & $ZWj$ & LO \\
$t \bar t$, $t \to q \gamma$ & $\gamma j \ell \nu b$ & $\gamma Wjj$ & LO$^{**}$
 & & $gq \to \gamma t$ & $\gamma \ell \nu b$ & $\gamma Wj$ & LO \\
$t \bar t$, $t \to q g$ & $j j \ell \nu b$ & $Wjjj$ & LO$^{*}$ &
 & $gq \to t$ & $\ell \nu b$ & $Wj$ & NLO$^{**}$ \\
$t \bar t$, $t \to q H$ & $b \bar b j \ell \nu b$ & $Wb \bar b jj$ & LO$^{*}$
 & & $gq \to Ht$ & $b \bar b \ell \nu b$ & $t \bar t$ & NLO$^{**}$ \\
\end{tabular}
\end{center}
\caption{Experimental signature and main background for several top rare decay
and single top production processes at LHC. The top antiquarks are assumed to
decay $\bar t \to W^- \bar b \to \ell^- \bar \nu_\ell \bar b$, and the $Z$ and
$H$ bosons in the channel $Z \to \ell^+ \ell^-$, $H \to b \bar b$.}
\label{tab:back}
\end{table}

In order to determine the discovery potential of these processes we
consider that only one FCN coupling $X_{qt}$, $\kappa_{qt}$, $\lambda_{qt}$,
$\zeta_{qt}$ or $g_{qt}$ is nonzero at a time. We give the limits for
$3 \, \sigma$ evidence, what happens when
the expected number of signal ($S$) plus SM background ($B$) events 
is not consistent with a background fluctuation at the $3 \,
\sigma$ level, corresponding to a CL of 0.9973. 
For large samples, this translates
into $S/\sqrt B = 3$, while for $B \leq 5$ events Poisson statistics must be
used.
We rescale the data in Refs.~\cite{han1,han2,gustavo,npb,hosch} to a common
$b$ tagging efficiency of 50\%
and a mistagging rate of 1\%, recalculating the limits using these unified
criteria.\footnote{In Ref.~\cite{han1} $b$ tagging is not used and to obtain our
limits we scale their cross sections by the appropriate factors. The
interactions considered there are of $\gamma_\mu$ type only but the limits for
$\sigma_{\mu \nu}$ couplings are expected to be the same. In
Ref.~\cite{han3} the analysis is done for Tevatron energies
only.} (We note that in these analyses a top quark mass $m_t \simeq 175$ GeV is
used.)
We assume an integrated luminosity of 100 fb$^{-1}$, corresponding to one year
of running in the high luminosity phase. 
For an increase in luminosity by a
factor $k$, the limits on branching ratios scale with $k^{-1/2}$.

We point out that in real experiments a proper consideration of theoretical
uncertainties in background cross sections will be compulsory.
Present calculations in the literature are aimed at determining the
sensitivity to FCN couplings of various processes, and do not need to take them
into account. However, for the
comparison of theoretical predictions with experimental data,
leading order (LO) background calculations will often be insufficient and 
next-to-leading order (NLO) calculations will be required 
to match the statistical precision achieved at LHC.
In Table~\ref{tab:back} we have written the order in perturbation theory to
which these backgrounds are presently known.
We estimate that when the statistical uncertainty of the
background cross sections\footnote{Including $b$ tagging and kinematical cuts,
and considering 100 fb$^{-1}$ of integrated luminosity. For a higher luminosity
the relative statistical uncertainty decreases.} is better than 20\% the use
of NLO calculations is necessary (this is indicated in Table~\ref{tab:back} by
an asterisk) and when it is
better than 5\%, next-to-next-to-leading-order calculations may be required
(indicated by a double asterisk).\footnote{In principle, it may be also
possible to normalise the background cross sections using measured data from
other phase space regions, thus decreasing the theoretical uncertainty in
the regions of interest. If this is the case, NLO or even
LO calculations may be sufficient.}

Our limits are collected in Table~\ref{tab:lim}.
In the majority of the cases top decay processes provide the best place to
discover top FCN
interactions, surpassed by single top production for $\sigma_{\mu \nu}$-type
interactions involving the up quark. Comparing these limits with the data in
Table~\ref{tab:br} we observe that in many examples the maximum rates predicted
are observable with $3 \, \sigma$ statistical significance or more within one
year (with a luminosity upgrade to 6000 fb$^{-1}$ \cite{LHCup} the figures in
Table~\ref{tab:lim} are reduced by a factor of 7.7).
If no signal is observed, upper bounds on top FCN decay
branching ratios can be placed. The 95\% upper limits obtained in this case
are numerically smaller than those in Table~\ref{tab:lim} by a factor
between 1.3 and 1.5.

\begin{table}[htb]
\vspace*{0.2cm}
\begin{center}
\begin{tabular}{lcc}
& Top decay & Single top \\
\hline\\[-0.5cm]
$t \to u Z (\gamma_\mu)$ & $3.6 \times 10^{-5}$ & $8.0 \times 10^{-5}$ \\
$t \to u Z (\sigma_{\mu \nu})$ & $3.6 \times 10^{-5}$ & $2.3 \times 10^{-5}$ \\
$t \to u \gamma$ & $1.2 \times 10^{-5}$ & $3.1 \times 10^{-6}$ \\
$t \to u g $ & $-$ & $2.5 \times 10^{-6}$ \\
$t \to u H $ & $5.8 \times 10^{-5}$ & $5.1 \times 10^{-4}$ \\
\end{tabular}
\quad
\begin{tabular}{lcc}
& Top decay & Single top \\
\hline\\[-0.5cm]
$t \to c Z (\gamma_\mu)$ & $3.6 \times 10^{-5}$ & $3.9 \times 10^{-4}$ \\
$t \to c Z (\sigma_{\mu \nu})$ & $3.6 \times 10^{-5}$ & $1.4 \times 10^{-4}$ \\
$t \to c \gamma$ & $1.2 \times 10^{-5}$ & $2.8 \times 10^{-5}$ \\
$t \to c g $ & $-$ & $1.6 \times 10^{-5}$ \\
$t \to c H $ & $5.8 \times 10^{-5}$ & $2.6 \times 10^{-3}$
\end{tabular}
\end{center}
\caption{$3 \, \sigma$ discovery limits for top FCN interactions at LHC, for an
integrated luminosity of 100 fb$^{-1}$. The limits are expressed in terms of top
decay branching ratios.}
\label{tab:lim}
\end{table}

The ATLAS and CMS collaborations have performed full detector simulations
to investigate
the sensitivity to the decays $t \to qZ$, $t \to q\gamma$, giving $5 \, \sigma$
discovery limits on the rates for these processes for an integrated luminosity
of 100 fb$^{-1}$. For the ATLAS detector the limits are $\mathrm{Br}(t \to qZ)
= 2.0 \times 10^{-4}$ \cite{atlas1}, $\mathrm{Br}(t \to q \gamma) = 1.0 \times
10^{-4}$ \cite{top}, and for the CMS detector $\mathrm{Br}(t \to qZ) = 1.9
\times 10^{-4}$, $\mathrm{Br}(t \to q \gamma) = 3.4 \times 10^{-5}$ \cite{top}.
After correcting for the different confidence levels used, the numbers for $t
\to q \gamma$ at CMS agree very well with those in Table~\ref{tab:lim}, while
the rest are more pessimistic.

To conclude this subsection we note that at LHC there are additional processes
which can occur through top FCN interactions. The first example is single top
production associated with a jet produced via $gtq$ interactions\cite{han4},
which is however less sensitive than $gq \to t$. The second example is
like-sign top production \cite{slabo}, mediated by two FCN vertices. This
process has a smaller cross section than processes with only one FCN vertex, but
might be observed at LHC due to its small background.

\subsection{Observation at an $\boldsymbol{e^+ e^-}$ collider}

\begin{figure}[htb]
\begin{center}
\vspace*{0,5cm}
\mbox{\epsfig{file=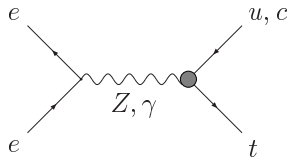,height=2cm,clip=}} 
\end{center}
\caption{Diagrams for single top production in $e^+ e^-$ collisions.}
\label{fig:eetq}
\end{figure}

A high energy $e^+ e^-$ collider like TESLA will complement LHC capabilities in
the search for top FCN couplings. As in hadron collisions, the operators in
Eq.~(\ref{ec:1})
mainly manifest themselves in top decay and single top production processes.
In $e^+ e^-$ annihilation top quark pairs are produced by electroweak
interactions, and single top quarks may be produced in the process
$e^+ e^- \to t \bar q$, \cite{hewett2}, via the diagrams in
Fig.~\ref{fig:eetq}. (The charge conjugate process is also summed.)
At TESLA the top pair production cross section at 500 GeV is only of 600 fb 
\cite{tesla}, and limits obtained from top decays \cite{mio2,sher}
cannot compete with those from LHC, despite
the larger luminosity and smaller backgrounds. On the contrary, single top
production can match or even improve some LHC limits if beam
polarisation is used to reduce backgrounds \cite{mio}. We have updated the
study of Ref.~\cite{mio} to include the effect of initial state radiation (ISR)
\cite{isr} and beamstrahlung \cite{BS2,peskin} in the calculations. We assume
integrated luminosities of 345 fb$^{-1}$ and 534 fb$^{-1}$ per year for centre
of mass (CM) energies of 500 and 800 GeV, respectively \cite{lum}, and beam
polarisations $P_{e^-} = 0.8$, $P_{e^+} = -0.6$.\footnote{Here we use the
convention in which the degree of polarisation refers to the helicity both for
the electron and the positron, in contrast with Refs.~\cite{mio,mio2}.}
For beamstrahlung at 500 GeV
we use the parameters $\Upsilon = 0.05$, $N = 1.56$,
while at 800 GeV we take $\Upsilon = 0.09$, $N = 1.51$ \cite{lum}. We also
include a beam energy spread of 1\%. The total
cross sections at both energies for each type of anomalous coupling are written
in Table~\ref{tab:cs2}.

\begin{table}[htb]
\begin{center}
\begin{tabular}{lcc}
& 500 GeV & 800 GeV \\
\hline \\[-0.5cm]
$Z,\gamma_\mu$ & $370 \, |X_{qt}|^2$ & $230 \, |X_{qt}|^2$ \\
$Z,\sigma_{\mu \nu}$ & $2560 \, |\kappa_{qt}|^2$ & $2850 \, |\kappa_{qt}|^2$ \\
$\gamma$ & $5370 \, |\lambda_{qt}|^2$ & $6300 \, |\lambda_{qt}|^2$ 
\end{tabular}
\end{center}
\caption{Cross sections (in fb) for single top production
at TESLA, including ISR, beamstrahlung and beam energy spread, for
polarisations $P_{e^-} = 0.8$, $P_{e^+} = -0.6$. The cross section for single
antitop production is the same.}
\label{tab:cs2}
\end{table}

We find that ISR and beamstrahlung make it more involved the reconstruction of
the top quark momentum and additionally they increase the $W\!jj$ background
cross
section. Following the analysis of 
Ref.~\cite{mio}, but with a different reconstruction procedure and different
sets of kinematical cuts, we obtain the $3 \, \sigma$ discovery limits in
Table~\ref{tab:lim_tesla}. The limits for $\gamma_\mu$ couplings to
the $Z$ boson are slightly better than the ones previously obtained in
Ref.~\cite{mio} without ISR and beamstrahlung, but still not competitive with
those from LHC. For $\sigma_{\mu \nu}$ interactions the opposite happens:
limits including these corrections are a little worse but at any rate
they improve the LHC potential in most cases, especially
at 800 GeV, where the $q^\nu$ factor in the vertex keeps signal cross sections
large.

\begin{table}[htb]
\begin{center}
\begin{tabular}{lcc}
& 500 GeV & 800 GeV \\
\hline\\[-0.5cm]
$t \to q Z (\gamma_\mu)$ & $1.9 \times 10^{-4}$ & $1.9 \times 10^{-4}$ \\
$t \to q Z (\sigma_{\mu \nu})$ & $1.8 \times 10^{-5}$ & $7.2 \times 10^{-6}$ \\
$t \to q \gamma$ & $1.0 \times 10^{-5}$ & $3.8 \times 10^{-6}$
\end{tabular}
\end{center}
\caption{$3 \, \sigma$ discovery limits for top FCN interactions in single top
production at TESLA, for CM energies of 500 and 800 GeV, with respective
luminosities of 345 fb$^{-1}$ and 534 fb$^{-1}$.
The limits are expressed in terms of top decay branching ratios.}
\label{tab:lim_tesla}
\end{table}

We remark that LHC and TESLA are complementary in the search for top FCN
interactions. LHC has a better discovery potential for $\gamma_\mu$ couplings
to the $Z$ boson and FCN interactions with the gluon and the Higgs boson, while
TESLA is more sensitive to $\sigma_{\mu \nu}$ couplings to the $Z$ and the
photon. Moreover, if positive signals are observed, results from both colliders
may be necessary to determine the type of operator involved. While TESLA
cannot disentangle $Z$ and photon interactions, its good $c$ tagging efficiency
may allow to determine the identity of the light quark. On the
contrary, the processes described at LHC determine if the FCN vertices involve
the $Z$ boson or the photon, but it is more difficult to tag the flavour of
the light quark.

\subsection{Other colliders}

For completeness, we list here other possible places where top FCN interactions
can be probed as well. One possibility is $e \gamma$ and $\gamma \gamma$
collisions. The latter is specially sensitive, and a positive
signal could be found in the context of the MSSM \cite{yang3,yang4}. Note
however that in this case
there are further contributions to $\gamma \gamma \to t \bar c$ given by box
diagrams which cannot be parameterised by the vertices in
$\mathcal{L}^\mathrm{eff}$. (This is also the case for $e^+ e^-$
annihilation \cite{wudka}.) $ep$ scattering is another place where this type
of interactions
might lead to new effects, but their sensitivity is far beyond the ones
achievable at LHC or a future $e^+ e^-$ collider.

\section{Conclusions}
\label{sec:5}

In the previous sections we have seen that top FCN couplings are negligible in
the SM but can be enhanced in SM extensions. We have shown that these
interactions lead to observable effects at high energy colliders, mainly in top
decay and single top production processes. In order to cleanly observe an excess
with respect to SM expectations, and hence the presence of top FCN
interactions, a precise background calculation is compulsory. This is a task
which should be carried out in the next few years, before LHC experimental data
are available.

We have shown that top FCN interactions offer a good place for the study of
indirect effects from physics beyond the SM. However, one important aspect which
has not been discussed is the correlation between top FCN processes and 
other possible new physics effects at high or low energies. 
This study includes, but is not limited to, the effect of top FCN operators in
low energy physics \cite{peris}. Although the branching
ratios in Table~\ref{tab:br} are in agreement with present experimental
data, effects in $B$ physics are possible and could be measured in experiments
under way at $B$ factories.

One example of such correlation is present in models with $Q=2/3$ singlets. A
coupling $|X_{ct}| \sim 0.015$ observable at LHC requires a sizeable deviation
of the diagonal $Ztt$ coupling from its SM expectation \cite{largo}, which would
certainly be seen in $t \bar t$ production at TESLA. Furthermore, a FCN coupling
of this size allows for a CP-violating phase 
$\chi = \arg (V_{ts} V_{tb}^* V_{cs}^* V_{cb}) \sim \pm 0.3$ in the CKM matrix
\cite{quico}, much larger in absolute value than the SM expectation
$0.015 \leq \chi \leq 0.022$. This phase would lead to observable phenomena in
$B$ oscillations and decay and, if such a phase is found, it necessarily
requires the presence of a FCN coupling at the observable level.

The examination of possible correlations between top FCN interactions and
other processes at low and high energies is model-dependent, and further
analyses should be done in that direction. In particular, if indirect effects
are meant to serve as consistency tests of a (new physics) model, the relation
between the presence of top FCN interactions at an observable level and other
indirect effects must be fully understood. The
investigation of such correlations will help uncover the nature of new
physics, if
positive signals are found at the present or next generation of colliders.

\vspace{1cm}
\noindent
{\Large \bf Acknowledgements}

\vspace{0.4cm} \noindent
I thank F. del Águila for discussions.
This work has been supported by the European Community's Human Potential
Programme under contract HTRN--CT--2000--00149 Physics at Colliders and by
FCT through projects CERN/FIS/43793/2002, CFIF--Plurianual (2/91)
and grant SFRH/BPD/12603/2003.

\end{document}